\documentclass[aps,showpacs,pre,superscriptaddress,twocolumn,nofootinbib]{revtex4}
\usepackage{amsfonts}
\usepackage{amsmath}
\usepackage{mathrsfs}
\usepackage{amssymb}
\usepackage{wasysym}
\usepackage{subfigure}
\usepackage{graphicx}% Include figure files
\usepackage{float}% Control the locations of figures [H]

\begin{document}
%\begin{CJK}{GBK}{song}

\title{Bogoliubov approach to superfluid-Bose glass phase transition of a disordered Bose-Hubbard Model in weakly interacting regime}
\author{Botao Wang}
\affiliation{Department of Physics, Shanghai University, Shanghai 200444, P.R. China}
\author{Ying Jiang}
\thanks{Corresponding author}
\email{yjiang@shu.edu.cn}
\affiliation{Department of Physics, Shanghai University, Shanghai 200444, P.R. China}
\affiliation{Shanghai Key Laboratory of High Temperature Superconductors, Shanghai 200444, P.R. China}

\begin{abstract}
We investigate the disorder effect on coherent fraction and the quantum phase transition of ultracold dilute Bose gases trapped in disordered optical lattices.
Within the framework of Bogoliubov theory, an analytical expression for the particle density is derived and the dependence of coherent fraction on disorder strength as well as on lattice depth is discussed. In weak disorder regime, we find a decreased sensitivity of coherent fraction to disorder with the increase of on-site interaction strength. For strong disorder, the quantum phase boundary between superfluid phase and Bose glass phase in the disordered Bose-Hubbard system in weak interaction regime is discussed qualitatively. The obtained phase diagram is in agreement with the empirical square-root law. The dependence of the corresponding critical value of the disorder strength on optical lattice depth is presented as well, and may serve as a reference object for possible experimental investigation.

%\vskip0.5cm
%\noindent
%$Key Words: Disordered Bose-Hubbard model, coherent fraction, quantum phase transition

\end{abstract}

\maketitle

\section{Introduction}

The appearance of ultracold quantum gases in optical lattices opens a new chapter in mimicking the condensed matter systems \cite{2007MLewenstein}.
Because of the unprecedented level of control and precision, such a so-called quantum simulator turns out to be an ideal tool in studying disordered systems \cite{2008IBloch,2015ODutta-review}.
Nowadays, it is routinely possible to create systems of ultracold atoms in different optical lattices \cite{2007MLewenstein,2008IBloch,2015ODutta-review,2015KVKrutisky} since the remarkable experimental work by Greiner \emph{et al.} \cite{2002MGreiner} in which the quantum phase transition from the superfluid (SF) to Mott insulator (MI) was first observed.
When disorder is introduced, there emerges a new phase, i.e. Bose glass (BG, quasi-superfluid puddles embedded in an MI insulating background) \cite{1989Fisher}.
Even though disorder, or pseudo-disorder can be realized in optical lattices in different ways \cite{2005JELye,2009MWhite,2012FJendrzejewski,2007LFallani,2011BGadway,2009MModugno}, directly observing such BG phase in disordered lattice systems has been a big challenge \cite{2007LFallani,2011BGadway,2014CDErrico,2010MPasienski}.
Only recently has it been reported that the phase transition of SF-BG can be detected in three-dimensional (3D) optical lattices by measuring the amount of excitations generated by quench \cite{2015CMeldgin-BG},
while the probe of SF-BG in 2D disordered optical lattices (to our knowledge) is still on the way.

Theoretically, a system with ultracold bosons trapped in optical lattices is often described by Bose-Hubbard model (BHM) with the disorder encrypted in the on-site potential \cite{1989Fisher,1998DJaksch}.
Hitherto, a variety of sophisticated methods have been applied in this study, including density-matrix renormalization group techniques, local mean-field approximation, stochastic mean field theory, strong coupling expansion,  quantum Monte Carlo simulation as well as  local mean-field cluster analysis (for references see Ref.[\onlinecite{2013AENiederle}] and references therein).
Most of the efforts, however, have been only devoted to the phase transition problem in strongly interacting regime, and little attention has been paid to a disordered Bose gas in lattice systems with weak interaction, where a phase transition from SF to BG can also occur when disorder strength is large enough \cite{2013AENiederle,2009VGurarie3DQMC,2011SGSoyler}.
Not only numerical methods face with a big challenge in properly revealing the phase boundary of SF-BG in weakly interacting regime because of the finite size effect \cite{2011SGSoyler,2015CZhang-QMC}, but a rigorous analytic investigation of the phase diagram in such a regime is still missing so far, except for an empirical square-root law estimation in 2D \cite{2011SGSoyler} that has been proposed in some way analogous to the disordered continuum (without optical lattices) case \cite{2009GMFalco-BG-SF}.
Thus, it is highly desirable to have a further exploration in the SF-BG phase transition of weakly-interacting Bose gases in disordered optical lattices.

In this work, we shall address this phase transition problem within the framework of Bogoliubov theory.
Considering its extensive applications and validity in weakly interacting systems \cite{1947NNBogoliubov,1992KHuang,2002GEAstrakharchik,2013GEAstrakharchik,2007GMFalco,2014CGaul,2014JSaliba,1994KGSingh,2013CGaul,2003AMRey,2006GOrso-BT,2009YHu-BT,2015MOCPires,2010PTErnst1-Ek,2011UBissbort-Ek,2006KXu-n-n0}, we extend Bogoliubov theory from the clean lattice system \cite{2001vOosten} to the disordered case and study the effect of disorder on the coherent fraction of dilute ultracold Bose gases (i.e. the fraction of particles with zero momentum \cite{2011GEAstrakharchik,2012CAMueller,2015CAMueller,2016ZDZhang} ).
We find that, in contrast to the clean system, Bogoliubov theory does have the potential to capture the disorder-induced SF-BG phase transition in weak interaction regime in disordered lattices. To this end we give a quantitative description of the dependence of coherent density on weak disorder and a qualitative picture of the SF-BG phase boundary in strong disorder regime.

The structure of the paper is as follows.
After introducing the disordered Bose-Hubbard Hamiltonian, we perform the Bogoliubov transformation and give the analytic expression of the particle density in Section II. In Section III, we investigate the disorder effect on coherent density and the SF-BG phase diagram by evaluating the dependence of coherent fraction on the the strength of disorder. In section IV we relate our theoretical results to the last 3D experimental data, and show our predictions in 2D systems, which is hoped to be a reference object for future experimental study in 2D. Finally, we summarize our results in section V.

\section{The Model and Bogoliubov Theory}

An ultracold dilute Bose gas in a disordered optical lattice can be depicted by the following disordered Bose-Hubbard Hamiltonian \cite{1989Fisher,1998DJaksch,2013AENiederle}
\begin{equation}
  \hat{H}_{BH}=-t{\displaystyle \sum_{\left\langle i,j\right\rangle }}\hat{a}_{i}^{\dagger}\hat{a}_{j}+\frac{U}{2}{\displaystyle \sum_{i}\hat{a}_{i}^{\dagger}\hat{a}_{i}^{\dagger}\hat{a}_{i}\hat{a}_{i}}-{\displaystyle \sum_{i}}\left(\mu+\epsilon_{i}\right)\hat{a}_{i}^{\dagger}\hat{a}_{i}
  \label{H-BH}
\end{equation}
where $\hat{a}_{i}$ and $\hat{a}^{\dagger}_{i}$ are bosonic annihilation and creation operators at site $i$, satisfying the canonical commutation.
$t$ is nearest-neighbor hopping parameter, $U$ represents the on-site repulsive interaction strength and $\mu$ is the chemical potential. $\epsilon_{i}$ describes the random potential, which is assumed to be uniformly distributed between $[-\Delta,\Delta]$ and spatially uncorrelated
\begin{align}
  \overline{\epsilon_{i}}=0,~~\overline{\epsilon_{i}\epsilon_{i^{\prime}}}=\frac{\Delta^{2}}{3}\delta_{i,i^{\prime}}
  \label{P-disorder}
\end{align}
where $\overline{\cdots}$ stands for the disorder ensemble average and $\Delta$ denotes the disorder strength.

After performing Fourier transformations for the operators $\hat{a}_{i}=\left(1/N_{S}\right)^{1/2}\sum_{\bf k}\hat{a}_{\bf k}e^{-{\bf i\cdot k}}$ (also for its corresponding conjugate) and the random potential $\epsilon_{i}=\left(1/N_{S}\right)^{1/2}\sum_{\bf k}{\epsilon_{{\bf k}}e^{-{\bf i\cdot k}}}$, the Hamiltonian Eq. (\ref{H-BH}) can be rewritten in momentum space as
\begin{align}
  \hat{H}= & -{\displaystyle \sum_{{\bf k}}}\left(2t{\displaystyle \sum_{l=1}^{d}}\cos ak_{l}+\mu\right)\hat{a}_{{\bf k}}^{\dagger}\hat{a}_{{\bf k}}
  \nonumber \\ &
  +\frac{U}{2}\frac{1}{N_{S}}{\displaystyle \sum_{{\bf k_{1},k_{2},k_{3},k_{4}}}}\hat{a}_{{\bf k_{1}}}^{\dagger}\hat{a}_{{\bf k_{2}}}^{\dagger}\hat{a}_{{\bf k_{3}}}\hat{a}_{{\bf k_{4}}}\delta_{{\bf k_{1}+k_{2},k_{3}+k_{4}}}
  \nonumber \\ &
  -\frac{1}{\sqrt{N_{S}}}{\displaystyle \sum_{{\bf k_{1},{\bf k_{2}},{\bf k_{3}}}}\epsilon_{{\bf k_{1}}}\hat{a}_{{\bf k_{2}}}^{\dagger}\hat{a}_{{\bf k_{3}}}}\delta_{{\bf k_{2},k_{1}+k_{3}}},
  \label{H-ak}
\end{align}
$N_{S}$ is the number of lattice sites, $a$ is the lattice constant, $d$ denotes the dimension and $\bf i$ represents the coordinate of site $i$.
Near absolute zero temperature, the number of atoms with zero momentum $N_{0}$ becomes macroscopically large, which allows the Bogoliubov prescription $\hat{a}_{{\bf 0}}\simeq\hat{a}_{{\bf 0}}^{\dagger}\simeq\sqrt{N_{{0}}}$ \cite{1947NNBogoliubov}.
In such a case, retaining all terms up to second order in $\hat{a}_{{\bf k}}^{\dagger}$, $\hat{a}_{{\bf k}}$ and $\epsilon_{{\bf k}}$ yields
\begin{align}
  \hat{H}= & \left(\frac{U}{2}n_{0}-2dt-\mu\right)N_{{\bf 0}}+{\displaystyle \sum_{{\bf k\neq0}}}A_{{\bf k}}\hat{a}_{{\bf k}}^{\dagger}\hat{a}_{{\bf k}}
  \nonumber \\ &
  +\frac{U}{2}n_{0}{\displaystyle \sum_{{\bf k\neq0}}}\left(\hat{a}_{{\bf k}}^{\dagger}\hat{a}_{{\bf -k}}^{\dagger}+\hat{a}_{{\bf k}}\hat{a}_{{\bf -k}}\right)
  \nonumber \\ &
  -\sqrt{n_{0}}{\displaystyle {\displaystyle \sum_{{\bf k\neq0}}}\left(\epsilon_{{\bf k}}\hat{a}_{{\bf k}}^{\dagger}+\epsilon_{{\bf -k}}\hat{a}_{{\bf k}}^{\dagger}\right)},
  \label{H-simplifed}
\end{align}
where we have introduced for brevity
\begin{align}
  A_{{\bf k}}=2Un_{0}-\mu-2t\sum_{l=1}^{d}\cos ak_{l}.
  \label{Ak}
\end{align}

\renewcommand{\thefootnote}{\fnsymbol{footnote}}
\setcounter{footnote}{0}

Note that the particle density with zero-momentum $n_{0}=N_{0}/N_{S}$ should be called coherent density \cite{2011GEAstrakharchik,2012CAMueller,2015CAMueller,2016ZDZhang}, which is linked with the long-range phase coherence of the whole system\footnotemark.
In inhomogeneous system, the condensate consists of coherent particles ($\bf k=0$) plus the glassy ones ($\bf k\neq0$, also named as deformed condensates) \cite{2012CAMueller,2013CGaul,2015CAMueller,2016ZDZhang}.
Thus the population of zero-momentum can not be used to determine the condensate fraction in systems in absence of translation invariance \cite{1956OPenrose,2011GEAstrakharchik,2012CAMueller,2015CAMueller,2016ZDZhang}.
The calculations of condensate fraction and condensate depletion can be performed via, for example, an inhomogeneous Bogoliubov theory developed by Gaul and Mueller \cite{2014CGaul,2013CGaul,2012CAMueller}.
In our present work, however, we focus on investigating the behavior of coherent particles by means of a different technique,  which in the end gives us some hints of the SF-BG phase transition in weakly-interacting regime.

\footnotetext[1]{The condensate fraction is defined as the largest eigenvalue of the one-body density matrix $\rho_{ij}=\left\langle \hat{a}_{i}^{\dagger}\hat{a}_{j}\right\rangle$ \cite{1956OPenrose,2003RRoth-Burnett,2007PBuonsante}, and the coherent density can be defined from the off-diagonal long-range order of the one-body density matrix, i.e. $n_{0}={\displaystyle \lim}_{\left|i-j\right|\rightarrow\infty}\rho_{ij}$ \cite{2012CAMueller,2015CAMueller,2011GEAstrakharchik,2016ZDZhang}. The difference between these two concepts in inhomogeneous systems was proposed by Penrose and Onsager \cite{1956OPenrose} and further clarified by M\"{u}ller and Gaul \cite{2012CAMueller,2015CAMueller}. See also \cite{2011GEAstrakharchik} for elaborate Monte Carlo simulations.}

The above Hamiltonian (\ref{H-simplifed}) can be diagonalized by the inhomogeneous Bogoliubov transformation \cite{1992KHuang,2007GMFalco}
\begin{equation}
  \begin{cases}
\hat{a}_{{\bf k}}^{\dagger}=u_{{\bf k}}\hat{b}_{{\bf k}}^{\dagger}-v_{{\bf k}}\hat{b}_{{\bf -k}}-z_{{\bf k}}\\
\hat{a}_{{\bf k}}=u_{{\bf k}}\hat{b}_{{\bf k}}-v_{{\bf k}}\hat{b}_{{\bf -k}}^{\dagger}-z_{{\bf k}}.
\label{a-b}
\end{cases}
\end{equation}
This transformation to new operators $\hat{b}_{{\bf k}}^{\dagger}$ ($\hat{b}_{{\bf k}}$) should preserve canonical commutation relations, which gives $u_{{\bf k}}^{2}-v_{{\bf k}}^{2}=1$.
A calculation along the standard procedure leads to the following Bogoliubov parameters
\begin{align}
  v_{{\bf k}}^{2}=& \frac{1}{2}\left(\frac{A_{{\bf k}}}{E_{{\bf k}}}-1\right),~
  u_{{\bf k}}^{2}=\frac{1}{2}\left(\frac{A_{{\bf k}}}{E_{{\bf k}}}+1\right),
  \\
  z_{{\bf k}}= & -\frac{\sqrt{n_{0}}}{A_{{\bf k}}+Un_{0}}\epsilon_{{\bf k}},
\end{align}
and the Bogoliubov spectrum
\begin{align}
  E_{{\bf k}}=\sqrt{A_{{\bf k}}^{2}-\left(Un_{0}\right)^{2}}.
  \label{Ek}
\end{align}
In the end, the diagonalized Hamiltonian reads
\begin{align}
  \hat{H}=& \left(\frac{Un_{0}}{2}-2dt-\mu\right)N_{0}+\frac{1}{2}{\displaystyle \sum_{{\bf k\neq0}}}\left(E_{{\bf k}}-A_{{\bf k}}\right)
  \nonumber \\ &
  -{\displaystyle \sum_{{\bf k\neq0}}}\frac{n_{0}}{A_{{\bf k}}+Un_{0}}\epsilon_{{\bf k}}^{2}+{\displaystyle \sum_{{\bf k\neq0}}}E_{{\bf k}}\hat{b}_{{\bf k}}^{\dagger}\hat{b}_{{\bf k}}.
  \label{H-diagonalized}
\end{align}

By now, the grand canonical-potential $\Omega$ can be obtained straightforwardly according to the definition $\Omega=-\beta^{-1}\ln\mathcal{Z}$, where $\mathcal{Z}= \rm{Tr} e^{-\beta\hat{H}}$ is the grand canonical partition function and $\beta=1/\left(k_{B}T\right)$ is the reciprocal temperature with $k_{B}$ being Boltzman constant. Since the grand canonical potential changes with each realization of disorder, we obtain the following thermodynamic potential by performing the disorder ensemble average
\begin{align}
  \overline{\Omega}= & \left(\frac{Un_{0}}{2}-\mu-2dt\right)n_{0}N_{s}+\frac{1}{2}{\displaystyle \sum_{{\bf k\neq0}}}\left(E_{{\bf k}}-A_{{\bf k}}\right)
  \nonumber \\ &
  -\frac{\Delta^{2}}{3}{\displaystyle \sum_{{\bf k\neq0}}}\frac{n_{0}}{A_{{\bf k}}+Un_{0}}+\frac{1}{\beta}{\displaystyle \sum_{{\bf k\neq0}}}\ln\left(1-e^{-\beta E_{{\bf k}}}\right)
  \label{Omega}
\end{align}
where $\overline{\epsilon_{{\bf k}}}=0,~\overline{\epsilon_{{\bf k}}^{2}}=\Delta^{2}/3$.
The above expression indicates that the grand canonical potential contains mean-field result (the first term on the right-hand side of the above expression), quantum fluctuation (the second term), contribution from disorder (the third term) and thermal fluctuation (the last term).

With the help of the thermodynamic relation $n=-\left(1/N_{S}\right)\partial\overline{\Omega}/\partial\mu$, we get a general expression of particle density in the framework of Bogoliubov theory
\begin{align}
     n= n_{0}+n_{I}+n_{T}+n_{R},
     \label{n-general}
\end{align}
which contains the coherent density $n_{0}$, the coherent depletion due to the on-site repulsive interaction
\begin{align}
n_{I}= \frac{1}{2}\frac{1}{N_{S}}{\displaystyle \sum_{{\bf k}\neq{\bf {\bf 0}}}}\left(\frac{A_{{\bf k}}}{E_{{\bf k}}}-1\right),
\end{align}
the temperature induced depletion
\begin{align}
n_{T}= \frac{1}{N_{S}}{\displaystyle \sum_{{\bf k}\neq{\bf {\bf 0}}}}\frac{1}{e^{\beta E_{{\bf k}}}-1}\frac{A_{{\bf k}}}{E_{{\bf k}}},
\label{n-T}
\end{align}
as well as the contribution coming from the random potential
\begin{align}
n_{R}= \frac{\Delta^{2}}{3}\frac{1}{N_{S}}{\displaystyle \sum_{{\bf k}\neq{\bf {\bf 0}}}}\frac{n_{0}}{\left(Un_{0}+A_{{\bf k}}\right)^{2}}.
\end{align}

From Eq. (\ref{Ak}), we see that the chemical potential $\mu$ is included in the expression of $A_{{\bf k}}$.
Hence, in order to go further to discuss the behavior of the particle density, the chemical potential needs to be determined first.
By minimizing the grand canonical potential $\overline{\Omega}$ in Eq. (\ref{Omega}) with respect to the coherent density $n_{0}$, i.e. $\left(1/N_{S}\right)\partial\overline{\Omega}/\partial n_{0}=0$, we have the chemical potential to the lowest order \cite{2001vOosten}
\begin{equation}
 \mu = Un_{0}-2dt,
 \label{mu-meanfield}
\end{equation}
which makes the Bogoliubov spectrum (\ref{Ek}) gapless in long-wavelength limit ${\bf k}\rightarrow{\bf 0}$, accordant with the Nambu-Goldstone theorem \cite{1960YNambu,1961JGoldstone}.
After taking the continuum limit \cite{2001vOosten}, we finally obtain the particle density expression of disordered lattice systems at zero-temperature
\begin{align}
  & n=n_{0}+n_{I}+n_{R},
  \label{n-q}
  \\
  & n_{I}=\frac{1}{2}{\displaystyle \prod_{l=1}^{d}}\int_{0}^{\pi/a}dk_{l}\left(\frac{a}{\pi}\right)\frac{Un_{0}+t_{k}}{\sqrt{t_{k}\left(2Un_{0}+t_{k}\right)}}-\frac{1}{2}
  ,\label{n-q-I}
  \\
  & n_{R}=\frac{\Delta^{2}}{3}{\displaystyle \prod_{l=1}^{d}}\int_{0}^{\pi/a}dk_{l}\left(\frac{a}{\pi}\right)\frac{n_{0}}{\left(2Un_{0}+t_{k}\right)^{2}}
  ,\label{n-q-R}
\end{align}
where we have introduced $t_{k}=2dt-2t\sum_{l=1}^{d}\cos ak_{l}$ for simplicity.
Note that when disorder is set to be zero ($\Delta=0$), our result exactly reduces to that obtained in clean lattice systems \cite{2001vOosten}.
Recently, the finite temperature effect in Eq. (\ref{n-T}) in clean cases has also been discussed \cite{2015MOCPires}.
In the following, we will focus on the disorder effect on the coherent density in the zero temperature limit.

Before some further discussions, we remark here that to arrive at Eq. (\ref{H-simplifed}), we have neglected the third and forth order terms of $\hat{a}_{{\bf k}}^{\dagger}$ ($\hat{a}_{{\bf k}}$), which represent high-order interactions of excited particles. Hence we will focus on the weakly-interacting regime where the on-site interaction is small enough for the Bogoliubov approximation to be valid (though large enough to make the one-band Bose-Hubbard model still work \cite{2003AMRey,2002GLAlfimov,2006OMorsch}).
At the same time, the disorder-related high-order terms $\epsilon_{{\bf k_{1}}}\hat{a}_{{\bf k_{2}}}^{\dagger}\hat{a}_{{\bf k_{3}}}$ with ${\bf k_{1},k_{2},k_{3}\neq 0}$ have also been discarded, so we expect that BT is rigorously valid in weak disorder region.
However, since there is only one term related to disorder among all the discarded high-order ones, there might be a possibility that the disorder strength may not need to be too weak.
Thus we would like to consider strong disorder as well.
In fact, the first attempt to study the properties of highly-disordered Bose condensates using Bogoliubov approximation has been made in a numerical way \cite{1994KGSingh}, where the condensate is shown to be destroyed for sufficiently large disorder and a BG phase may be reached.
In this paper, we also push the Bogoliubov approximation to its strong-disorder limit, but by investigating the coherent fraction analytically, with the hope to obtain some qualitative results representative of such a disorder-driven SF-BG phase transition in weakly-interacting regime.
Further improvement may need to take into account all the discarded higher-order terms in Hamiltonian (\ref{H-simplifed}), in which case the validity of the Bogoliubov approximation could be investigated rigorously.

\section{Disorder effect on coherent fraction}

\begin{figure}[H]
\centering\includegraphics[width=0.8\linewidth]{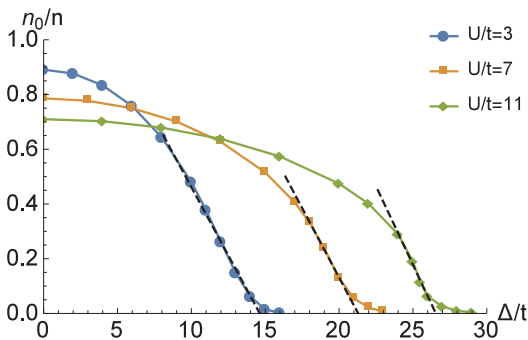}
\caption{(Color online) Coherent fraction $n_{0}/n$ versus disorder strength $\Delta/t$ for different interaction strengths $U/t$ in 2D. Dashed lines show the linear fits for fast decreasing parts of the curves, and are used to determine the transition points.}
\label{n0-Delta}
\end{figure}

By treating Eqs. (\ref{n-q}), (\ref{n-q-I}) and (\ref{n-q-R}) as a set of equations, the dependence of the coherent fraction $n_{0}/n$ on disorder strength $\Delta/t$ for different interaction strengths $U/t$ can be determined rigorously via implicit function technique. The main results in 2D ($d=2$) are shown in Fig. \ref{n0-Delta}. From the picture we see that, in the absence of disorder, $n_{0}/n$ is less than unity because mutual interaction will scatter the particles out of the coherent state. The scattering effect is getting more intense for stronger interaction strength, i.e. the coherent fraction is smaller for larger $U/t$.
When disorder is introduced and keeps at small values, there is only a very slight decrease in $n_{0}/n$, which implies that weak disorder hardly affects the coherent fraction.

In the intermediate disorder region, Fig. \ref{n0-Delta} exhibits a crossing behavior for different values of interaction strength $U/t$. This phenomenon can be explained by that the system becomes more robust to the addition of disorder with the increase of $U/t$, i.e. larger disorder is required to deplete the same amount of coherent fraction. This decreased sensitivity of $n_{0}/n$ to disorder with increasing interaction can be regarded as a criterion for superfluidity \cite{1994KGSingh}, so the systems remains in SF phase within this disorder range.
Note that this insensitivity phenomenon happens in the region with small coherent depletion (less than 50$\%$ ), where the Bogoliubov approximation is valid. Such a crossing behavior can also be found in disordered systems without lattice potential \cite{2013GEAstrakharchik}.

\begin{figure}[H]
\centering\includegraphics[width=\linewidth]{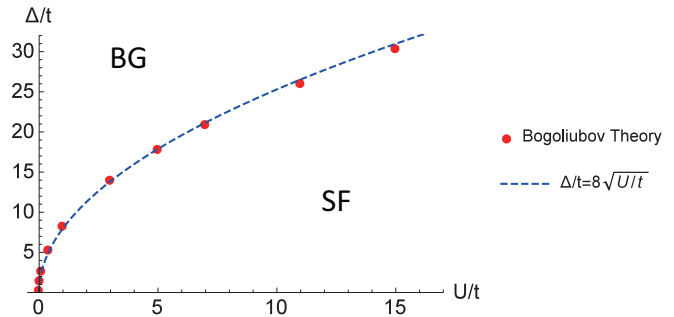}~~~~
\caption{(Color online) Zero temperature phase diagram of BG-SF phase transition in $\Delta/t$-$U/t$ plane obtained analytically by the use of Bogoliubov theory in the weakly interacting regime of 2D disordered BHM.}
\label{SF-BG-1}
\end{figure}

For strong disorder, $n_{0}/n$ goes down at a relatively high speed.
When disorder becomes sufficiently large, the decreasing of $n_{0}/n$ dramatically slows down, leading to the long tail structures shown in Fig. \ref{n0-Delta}.
As coherent depletion (with $\bf k\neq0$) becomes large, the third and fourth order terms discarded in Hamiltonian (\ref{H-simplifed}) will play a more and more important role. Thus the Bogoliubov approximation becomes worse and our calculations in the strong-disorder regime can not be considered as quantitative.
Nevertheless, the graphs in Fig. \ref{n0-Delta} may allow us to give a qualitative investigation. As displayed in the figure, $n_{0}/n$ approaches to zero for large enough $\Delta/t$, which corresponds to the loss of global phase coherence over the whole system.
Thus we assume that this threshold behavior in disorder strength can be associated with the entrance into BG phase from SF phase.
To approximately estimate the corresponding critical points $\Delta_{c}/t$ for different values of $U/t$, we look for the intersection points of the dashed lines and the asymptotes of the long tails of the curves (which are the horizontal axis) in Fig. \ref{n0-Delta}.
Such a procedure \footnotemark[2] leads to the phase diagram shown in Fig. \ref{SF-BG-1}, which agrees with the empirical square-root law $\Delta/t\propto\sqrt{U/t}$ \cite{2011SGSoyler,2009GMFalco-BG-SF}, and the coefficient is estimated to be 8 in our result.

\footnotetext[2]{Remarkably, while applying such a process to the experimental data measured in 3D disordered optical lattice system \cite{2010MPasienski}, one can find that the obtained critical point ($\Delta/E_{R}=3$ also corresponds to $s\simeq12$) ) agrees quite well with that gotten from quantum Monte Carlo simulation \cite{2009VGurarie3DQMC}}

\iffalse
Although Bogoliubov approximation may break down for strong disorder, its predictions in this regime could be qualitatively correct \cite{2002GEAstrakharchik,2013GEAstrakharchik}.
Besides the accordance with square-root law mentioned above, the long tail structures in figure \ref{3Dn0-Delta} could even be explained by the possible existence of a BG phase where the coherent fraction stays at very small values.
In fact, recent diffusion Monte Carlo simulations does suggest an insulating state with vanishing superfluid density but finite coherent fraction in a continuum model \cite{2013GEAstrakharchik}.
In 2D disordered optical lattices, the Gutzwiller approach has predicted that in BG phase, the superfluid fraction vanishes while the condensate fraction tends to a small value (about $2\%$) \cite{2003BDamski0.02}.
A similar observation can also be made according to a Monte Carlo study of SF-BG phase transition in 3D disordered Bose-Hubbard model \cite{2006PHitchcock0.025}.
On the experiment side, Bogoliubov theory turns out to provide a semiquantitative description for the measurements of quantum depletion of Bose-Einstein condensates trapped in optical lattices with the depletion fraction even in excess of $50\%$ \cite{2006KXu-n-n0}.
Thus we link our theoretical predictions to possible experiment realizations in the next section.
\fi

\section{Link theory to experiment}

Thanks to the time-of-flight technique, the dependence of coherent depletion on optical lattice depth can be directly measured in experiments \cite{2010MPasienski,2006KXu-n-n0,2003WZwerger}.
Meanwhile, it has been reported recently that Meldgin \emph{et al.} have successfully probed the SF-BG transition of 3D disordered BHM \cite{2015CMeldgin-BG} and the corresponding disorder strength $-$ lattice depth phase diagram has been obtained. In the following, we express our results in terms of experimental parameters to make a comparison with the present experimental data.

As is known, under a single-band tight binding approximation, the Bose-Hubbard Hamiltonian (\ref{H-BH}) can be derived from a disordered many-body Hamiltonian \cite{1998DJaksch}.
Exact numerical results for the on-site interaction strength can be accurately fitted by \cite{2015KVKrutisky}
\begin{equation}
  \frac{U}{E_{R}}=\frac{8}{\pi}\left(\frac{\pi}{4}\frac{\hbar\omega_{\bot}}{E_{R}}\right)^{\frac{3-d}{2}}\left[u\left(\frac{V_{0}}{E_{R}}\right)\right]^{\frac{d}{3}}\frac{a_{s}}{a},
  \label{U-ER}
\end{equation}
where $u\left(x\right)=\sum_{i=0}^{6}p_{i}x^{i}$ is a polynomial function with the coefficients $p_{0}=8/27, ~p_{1}=0.554092, ~p_{2}=0.0801432, ~p_{3}=-8.94513\times10^{-3}, ~p_{4}=4.55577\times10^{-4}, ~p_{5}=-1.12896\times10^{-5},~ p_{6}=1.09512\times10^{-7}$. The expression of hopping amplitude takes the form \cite{2015KVKrutisky}
\begin{align}
  \frac{t}{E_{R}}= & q_{1}\left(\frac{V_{0}}{E_{R}}\right)^{q_{2}}\exp\left[-q_{3}\left(\frac{V_{0}}{E_{R}}\right)^{q_{4}}\right]
  \nonumber \\ & +\frac{2}{\pi^{2}}\exp\left[-q_{5}\left(\frac{V_{0}}{E_{R}}\right)^{q_{6}}\right]
  \label{t-ER}
\end{align}
with $q_{1}=0.116828, ~q_{2}=1.16938, ~q_{3}=1.11717, ~q_{4}=0.63, ~q_{5}=0.369658, ~ q_{6}=1.01448$.
Here, $V_0$ represents the optical lattice depth in horizontal plane and $a=\lambda/2$ is the lattice constant with $\lambda$ being the wavelength of the laser. $\omega_{\bot}$ is the confinement frequency in third dimension. $a_s$ is the $s$-wave scattering length and $E_{R}=2\hbar^{2}\pi^{2}/m\lambda^{2}$ is the recoil energy.

\begin{figure}[H]
\centering\includegraphics[width=\linewidth]{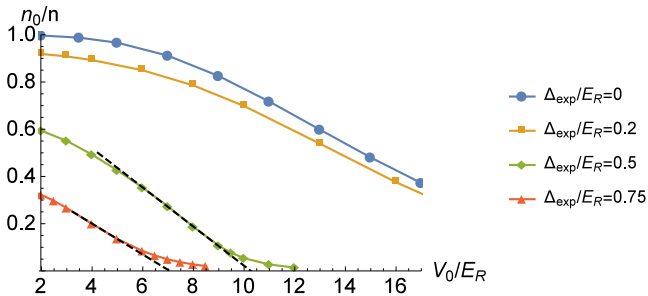}
\caption{(Color online) Coherent fraction $n_{0}/n$ versus optical lattice strength $V_{0}/E_{R}$ for different disorder strength $\Delta_{exp}/E_{R}$ in 3D. Dashed lines show the linear fits used to determine the transition points.}
\label{3Dn0-U}
\end{figure}

Combining the particle density expressions Eqs. (\ref{n-q}), (\ref{n-q-I}), and (\ref{n-q-R}) with Eqs. (\ref{U-ER}), (\ref{t-ER}), in Fig. \ref{3Dn0-U} we show the relationship between coherent fraction $n_0/n$ and optical lattice depth $V_0/E_R$ for different disorder strength $\Delta_{exp}$ in 3D. Here $\Delta_{exp}=\Delta/\sqrt{3}$ is the standard derivation of the distribution of site occupation energies which is commonly adopted in experiments.
From this figure we see that although coherent fraction always remains at some finite value for weak disorder strength (where the on-site interaction is still dominant), it will approach to zero when the strength of disorder is strong enough, which qualitatively agrees with previous experimental observations in 3D disordered optical lattice systems \cite{2010MPasienski}.

\begin{figure}[H]
\centering\includegraphics[width=\linewidth]{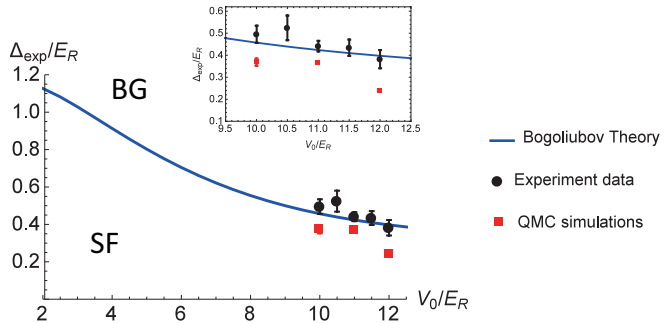}
\caption{(Color online) Zero temperature phase diagram of BG-SF obtained by Bogoliubov theory (Blue line) in the weakly interacting regime of 3D disordered BHM in terms of experimental parameters. The black dots and red squares are the mostly recent experimental results and quantum Monte Carlo simulations respectively \cite{2015CMeldgin-BG}.}
\label{3DSF-BG}
\end{figure}

Along the same avenue of obtaining Fig. \ref{SF-BG-1}, based on the result shown in Fig. \ref{3Dn0-U}, the $\Delta_{exp}/E_{R}-V_{0}/E_{R}$ phase diagram of SF-BG phase transition for the 3D system can be given, as shown in Fig. \ref{3DSF-BG}. Surprisingly, a very good agreement can be observed while comparing our theoretical prediction with the last experimental and numerical results. The red squares come from quantum Monte Carlo simulations and the black dots are obtained from the measurements of excitations produced by quantum quenches of disorder, where a threshold behavior in the disorder strength is associated with the phase transition \cite{2015CMeldgin-BG}.
Although this quantitative accordance might be a coincidence, we argue that Bogoliubov theory at least has the ability to correctly capture the qualitative feature of SF-BG phase transition in weakly interacting regime.

Nowadays, probing the SF-BG phase transition in 2D optical lattices has not been realized yet. Thus it would be helpful to find out the dependence of coherent fraction on lattice depth for different disorder strengths as well as the corresponding phase diagram in disorder $-$ lattice depth plane for 2D systems.

\begin{figure}[H]
\centering\includegraphics[width=\linewidth]{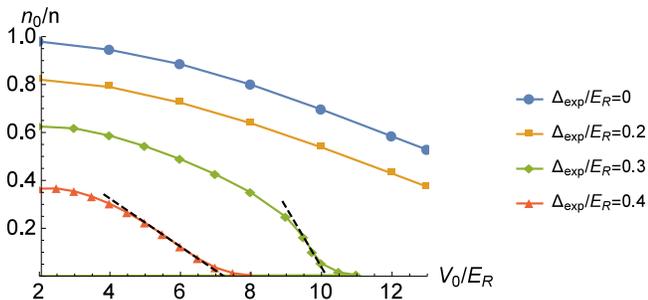}
\caption{(Color online) Coherent fraction $n_{0}/n$ versus optical lattice strength $V_{0}/E_{R}$ for different disorder strength $\Delta/E_{R}$. Dashed lines show the linear fits used to determine the transition points.}
\label{n0-U}
\end{figure}

\begin{figure}[H]
\centering\includegraphics[width=0.9\linewidth]{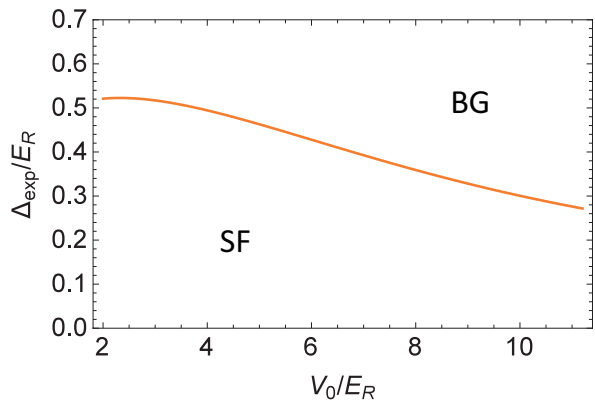}
\caption{(Color online) Zero temperature phase diagram of BG-SF obtained by Bogoliubov theory in the weakly interacting regime of 2D disordered BHM in terms of experimental parameters.}
\label{SF-BG}
\end{figure}

To plot the figure, we take the experimental parameters of a 2D optical lattice systems from Ref. \cite{2010CLHung-CChin}, where a Bose gas of $^{133}$Cs atoms is loaded and $\lambda=1064nm$, $\omega_{\bot}=1970\cdot2\pi Hz$, $a_{s}=200a_{B}$ ($ a_{B}$ being the Bohr radius).
In this case, combining Eqs. (\ref{n-q}), (\ref{n-q-I}), and (\ref{n-q-R}) with Eqs. (\ref{U-ER}), (\ref{t-ER}), we get the relationship between coherent fraction $n_0/n$ and optical lattice depth $V_0/E_R$ in 2D case in Fig. \ref{n0-U} , which turns out to be in qualitative agreement with that in 3D.
Along the similar way, the $\Delta/E_{R}-V_{0}/E_{R}$ phase diagram of SF-BG phase transition for the 2D system is also determined in Fig. \ref{SF-BG}. We find that the critical disorder strength $\Delta_c/E_{R}$ decreases with the increase of lattice depth $V_{0}/E_{R}$, which is in accordance with the experiment observations in 3D case \cite{2015CMeldgin-BG}.
Our results are expected to serve as a reference object for further studies of 2D disordered Bose systems in optical lattices.
%Meanwhile, while comparing the graph in Fig. \ref{SF-BG} with that in Fig. \ref{3DSF-BG}, one can easily find that at a given optical depth, the SF-BG phase transition happens at a smaller critical disorder strength in 2D than that in 3D.

\section{Summery}

In this paper we have shown that Bogoliubov theory turns out to be a simple yet effective approach in studying dilute weakly interacting Bose gases in 2D optical lattices with spatially uncorrelated disorder.
Under the Bogoliubov approximation, we have obtained the analytic expression of particle density in disordered lattice systems. When setting disorder strength to be zero, our result exactly goes back to the former calculation gotten in the clean case \cite{2001vOosten}.
By analyzing the implicit expressions of coherent fraction $n_{0}/n$, we have obtained the relationships between $n_{0}/n$ and disorder strength as well as lattice depth.
In weak disorder regime, Bogoliubov theory correctly capture the disorder effect on $n_{0}/n$, i.e. weak disorder hardly depletes the coherent fraction and the sensitivity of $n_{0}/n$ to disorder decreases as the interaction strength $U/t$ grows.

At large values of disorder, a qualitative picture of the SF-BG phase transition of weakly-interacting Bose gases has been given. The coherent fraction will approach to zero for sufficiently large disorder strength (although Bogoliubov approximation breaks down in this regime), which implies the loss of long-range phase correlation and is associated with the SF-BG phase transition. The zero temperature phase boundary of SF-BG is obtained. Our result turns out to be in agreement with the empirical square-root law \cite{2009GMFalco-BG-SF,2011SGSoyler}, and could be examined experimentally in the near future.

\begin{acknowledgments}
The authors greatly acknowledge Axel Pelster for stimulating and fruitful
discussions. We are also very grateful to C. Gaul and C. A. M\"{u}ller for their elaborate clarifications of the differences between coherent fraction and condensate fraction. This Work was supported by National Natural Science Foundation of China under Grant No. 11275119 and by Ph.D. Programs Foundation of Ministry of Education of China under Grant No. 20123108110004. Support from Shanghai Key Laboratory of High Temperature Superconductors (No. 14DZ2260700) is also acknowledged.
\end{acknowledgments}

%\end{CJK}

\begin{thebibliography}{46}

%Review
\bibitem{2007MLewenstein} M. Lewenstein, A. Sanpera and V. Ahufinger \emph{Ultracold Atoms in Optical Lattices, Simulating Quantum Many-Body Systems} (Oxford: Oxford University Press, 2012).
\bibitem{2015ODutta-review} O. Dutta, M. Gajda, P. Hauke, M. Lewenstein, D. S. L\"{u}hmann, B. A. Malomed, T. Sowi¨½ski and J. Zakrzewski, Rep. Prog. Phys. \textbf{78}, 066001(2015).
\bibitem{2008IBloch} I. Bloch, J. Dalibard and W. Zwerger, Rev. Mod. Phys. \textbf{80}, 885 (2008).
\bibitem{2015KVKrutisky} K. V. Krutisky, Phys. Rep. \textbf{607}, 1-101 (2016).
%Experiments


\bibitem{2002MGreiner} M. Greiner, O. Mandel, T. Esslinger, T. W. H\"{a}nsch and I. Bloch, Nature \textbf{415}, 39 (2002).
\bibitem{1989Fisher} M. P. A. Fisher, P. B. Weichman, G. Grinstein and D. S. Fisher, Phys. Rev. B \textbf{40}, 546 (1989).

\bibitem{2005JELye} J. E. Lye, L. Fallani, M. Modugno, D. S. Wiersma, C. Fort and M. Inguscio, Phys. Rev. Lett. \textbf{95}, 070401 (2005).
\bibitem{2009MWhite} M. White, M. Pasienski, D. McKay, S. Q. Zhou, D. Ceperley, and B. DeMarco, Phys. Rev. Lett. \textbf{102}, 055301 (2009).
\bibitem{2012FJendrzejewski} F. Jendrzejewski, A. Bernard, K. Muller, P. Cheinet, V. Josse, M. Piraud, L. Pezze, L. Sanchez-Palencia, A. Aspect and P. Bouyer, Nat. Phys. \textbf{8}, 398 (2012).
\bibitem{2007LFallani} L. Fallani, J. E. Lye, V. Guarrera, C. Fort and M. Inguscio, Phys. Rev. Lett. \textbf{98}, 130404 (2007)
\bibitem{2009MModugno} M. Modugno, New J. Phys. \textbf{11}, 033023 (2009)


\bibitem{2011BGadway} B. Gadway, D. Pertot, J. Reeves, M. Vogt and D. Schneble, Phys. Rev. Lett. \textbf{107}, 145306 (2011).
\bibitem{2014CDErrico} C. D'Errico, E. Lucioni, L. Tanzi, L. Gori, G. Roux, I. P. McCulloch, T. Giamarchi, M. Inguscio and G. Modugno, Phys. Rev. Lett. \textbf{113}, 095301 (2014).
\bibitem{2010MPasienski} M. Pasienski, D. McKay, M. White and B. DeMarco, Nat. Phys. \textbf{6}, 677 (2010).
\bibitem{2015CMeldgin-BG} C. Meldgin, U. Ray, P. Russ, D. Ceperley and B. DeMarco, arXiv: 1502. 02333v1.

\bibitem{1998DJaksch} D. Jaksch, C. Bruder, J. I. Cirac, C. W. Gardiner and P. Zoller, Phys. Rev. Lett. \textbf{81}, 3108 (1998).



%QMC
\bibitem{2013AENiederle} A. E. Niederle and H. Rieger, New J. Phys. \textbf{15}, 075029 (2013).
\bibitem{2009VGurarie3DQMC} V. Gurarie, L. Pollet, N. V. Prokof¡¯ev, B. V. Svistunov and M. Troyer, Phys. Rev. B \textbf{80}, 214519 (2009).
\bibitem{2011SGSoyler} S. G. S\"{o}yler, M. Kiselev, N. V. Prokof¡¯ev and B. V. Svistunov, Phys. Rev. Lett. \textbf{107}, 185301 (2011).
\bibitem{2015CZhang-QMC}C. Zhang, A. Safavi-Naini and B. Capogrosso-Sansone, Phys. Rev. A \textbf{91}, 031604(R) (2015)
\bibitem{2009GMFalco-BG-SF} G. M. Falco, T. Nattermann and V. L. Pokrovsky, Europhys. Lett. \textbf{85}, 30002 (2009); Phys. Rev. B \textbf{80}, 104515 (2009).


%Bogoliubov - continue

\bibitem{1947NNBogoliubov} N. N. Bogoliubov, Izv. Acad. Nauk (USSR) \textbf{11}, 77 (1947) [J. Phys. \textbf{11}, 23 (1947)].
\bibitem{1992KHuang} K. Huang and H. F. Meng, Phys. Rev. Lett. \textbf{69}, 644 (1992).
\bibitem{2002GEAstrakharchik} G. E. Astrakharchik, J. Boronat, J. Casulleras and S. Giorgini, Phys. Rev. A \textbf{66}, 023603 (2002).
\bibitem{2013GEAstrakharchik} G. E. Astrakharchik, K. V. Krutitsky and P. Navez, Phys. Rev. A \textbf{87}, 061601(R) (2013).
\bibitem{2007GMFalco} G. M. Falco, A. Pelster and R. Graham, Phys. Rev. A \textbf{75}, 063619 (2007); M. Ghabour and A. Pelster, Phys. Rev. A \textbf{90}, 063636 (2014).
\bibitem{2014CGaul} C. Gaul and C. A. M\"{u}ller, Appl. Phys. B \textbf{117}, 775¨C784 (2014).
\bibitem{2014JSaliba} J. Saliba, P. Lugan and V. Savona, Phys. Rev. A \textbf{90}, 031603(R) (2014).

%Bogoliubov - lattice
\bibitem{2003AMRey} A. M. Rey, K. Burnett, R. Roth, M. Edwards, C. J. Williams and C. W. Clark, J. Phys. B: At. Mol. Opt. Phys. \textbf{36}, 825 (2003).
\bibitem{2006GOrso-BT} G. Orso, C. Menotti and S. Stringari, Phys. Rev. Lett. \textbf{97}, 190408 (2006).
\bibitem{2009YHu-BT} Y. Hu, Z. X. Liang and B. B. Hu, Phys. Rev. A \textbf{80}, 043629 (2009).
\bibitem{1994KGSingh} K. G. Singh and D. S. Rokhsar, Phys. Rev. B \textbf{49}, 9013 (1994)
\bibitem{2013CGaul} C. Gaul and C. A. M\"{u}ller, Eur. Phys. J. Special Topics \textbf{217}, 69 (2013).
\bibitem{2015MOCPires} M. O. C. Pires and E. J. V. de Passos, arXiv: 1505. 01875v1.


% check BT
\bibitem{2010PTErnst1-Ek} P. T. Ernst1, S. G\"{o}tze1, J. S. Krauser1, K. Pyka1, D. S. L\"{u}hmann, D. Pfannkuche and K. Sengstock, Nat. Phys. \textbf{6}, 56 (2010).
\bibitem{2011UBissbort-Ek} U. Bissbort, S. G\"{o}tze, Y. Li, J. Heinze, J. S. Krauser, M. Weinberg, C. Becker, K. Sengstock and W. Hofstetter, Phys. Rev. Lett. \textbf{106}, 205303 (2011).
\bibitem{2006KXu-n-n0} K. Xu, Y. Liu, D. E. Miller, J. K. Chin, W. Setiawan and W. Ketterle, Phys. Rev. Lett. \textbf{96}, 180405 (2006).

\bibitem{2001vOosten}  D. van Oosten,  P. van der Straten and  H. T. C. Stoof, Phys. Rev. A \textbf{63}, 053601 (2001).

%coherent fraction

\bibitem{2011GEAstrakharchik} G. E. Astrakharchik and K. V. Krutitsky, Phys. Rev. A \textbf{84}, 031604(R) (2011).
\bibitem{2012CAMueller} C. A. Mueller and C. Gaul, New J. Phys. \textbf{14}, 075025 (2012).
\bibitem{2015CAMueller} C. A. Mueller, Phys. Rev. A \textbf{91}, 023602 (2015).
\bibitem{2016ZDZhang} L. Chen, Z. X. Liang, Y. Hu and Z. D. Zhang, J. Phys. B: At. Mol. Opt. Phys. \textbf{49}, 025303 (2016).

%\bibitem{2002AVLopatin} A. V. Lopatin and V. M. Vinokur, Phys. Rev. Lett. \textbf{88}, 235503 (2002).

\bibitem{1956OPenrose} O. Penrose and L. Onsager, Phys. Rev. \textbf{104}, 576 (1956).
\bibitem{2003RRoth-Burnett} R. Roth and K. Burnett, Phys. Rev. A \textbf{67}, 031602(R) (2003).
\bibitem{2007PBuonsante} P. Buonsante, V. Penna, A. Vezzani and P. B. Blakie, Phys. Rev. A \textbf{76}, 011602(R) (2007).

%finite n0
%\bibitem{2003BDamski0.02} B. Damski, J. Zakrzewski, L. Santos, P. Zoller and M. Lewenstein, Phys. Rev. Lett. \textbf{91}, 080403 (2003).
%\bibitem{2006PHitchcock0.025} P. Hitchcock and E. S. S{\o}rensen, Phys. Rev. B \textbf{73}, 174523 (2006).


%gapless spectrum
\bibitem{1960YNambu}  Y. Nambu, Phys. Rev. Lett. \textbf{4}, 380 (1960).
\bibitem{1961JGoldstone} J. Goldstone, Nuovo Cimento \textbf{19}, 154 (1961).

%BHM--V0

\bibitem{2002GLAlfimov} G. L. Alfimov, P. G. Kevrekidis, V. V. Konotop and M. Salerno, Phys. Rev. E \textbf{66}, 046608 (2002).
\bibitem{2006OMorsch} O. Morsch and M. Oberthaler, Rev. Mod. Phys. \textbf{78}, 179 (2006)


\bibitem{2003WZwerger} W. Zwerger, J. of Opt. B: Quantum and Semiclass. Opt. \textbf{5}, S9 (2003).

%qusi-2D
%\bibitem{2009NGemelke-CChin} N. Gemelke, X. Zhang, C. L. Hung and C. Chin, Nature \textbf{460}, 995 (2009).
\bibitem{2010CLHung-CChin} C. L. Hung, X. Zhang, N. Gemelke and C. Chin, Phys. Rev. Lett. \textbf{104}, 160403 (2010).


\end{thebibliography}
\end{document}